\documentclass[prc,twocolumn,showpacs]{revtex4}

\usepackage{graphicx}
\usepackage{dcolumn}
\usepackage{bm}
\usepackage{enumerate}

\usepackage{amsmath}
\usepackage{amssymb}
\usepackage{Ulem}
\usepackage{times}

\newcommand{\beq}{\begin{equation}}
\newcommand{\eeq}{\end{equation}}
\newcommand{\bea}{\begin{eqnarray}}
\newcommand{\eea}{\end{eqnarray}}

\def\la{\mathrel{\mathpalette\fun <}}
\def\ga{\mathrel{\mathpalette\fun >}}
\def\fun#1#2{\lower3.6pt\vbox{\baselineskip0pt\lineskip.9pt
 \ialign{$\mathsurround=0pt#1\hfil##\hfil$\crcr#2\crcr\sim\crcr}}}


\begin{document}

\title{
Self-consistent
microscopic description of neutron scattering by $^{16}$O
\\ based on the continuum particle-vibration coupling method}

\author{Kazuhito Mizuyama}
\email{mizukazu147@gmail.com}
\author{Kazuyuki Ogata}

\affiliation{
Research Center for Nuclear Physics, Osaka University, Ibaraki 567-0047, Japan
}

\date{\today}

\begin{abstract}
The microscopic description of neutron scattering by $^{16}$O below 30~MeV
is carried out by means of the continuum particle-vibration coupling (cPVC) method with the Skyrme nucleon-nucleon ($NN$)
effective interaction.
In the cPVC method, a proper boundary condition on a nucleon in continuum states
is imposed, which enables one to evaluate the transition matrix in a
straightforward manner.
Experimental data of
the total and total-elastic cross sections are
reproduced quite well by the cPVC method.
An important feature of the result
is the fragmentation of the single-particle resonance into
many peaks as well as the shift of its centroid energy. Thus,
some part of the fine structure of the experimental cross sections
at lower energies is well described by the cPVC framework.
The cPVC method based on a real $NN$ effective interaction
is found to successfully explain about 85\% of the reaction cross
section, through explicit channel-coupling effects.
\end{abstract}

\maketitle

Description of nucleon-nucleus ($NA$) elastic
scattering based on the fundamental nucleon-nucleon ($NN$) interaction
is one of the most challenging subjects of nuclear reaction studies,
and is crucial for exploration of unstable nuclei, for which
phenomenological optical potentials have not been established.
The most essential quantity for this subject is the imaginary part $W$
of the optical potential,
which is responsible for
a loss of the incident flux due to the existence
of non-elastic channels.
Reliability of $W$ can be judged by
comparing the resulting reaction cross section $\sigma_{\rm R}$
with experimental data.

One of the most successful approaches to this goal is the folding
model based on a complex effective $NN$ interaction and a phenomenological
or microscopic nuclear density. In Ref.~\cite{Deb05}, for example,
the folding model calculation with no free parameter is shown to
well reproduce differential cross sections and spin observables
for proton elastic scattering on
$^{12}$C, $^{16}$O, $^{40}$Ca, $^{90}$Zr, and $^{208}$Pb
at 65--200~MeV, as well as $\sigma_{\rm R}$ of neutron on these targets
at 20--800~MeV.
The agreement is at almost the same level as that of the
well-established Dirac phenomenology~\cite{Ray92,Coo93}. Many studies
in this direction have successfully been done with the coordinate-space
representation~\cite{Rik84,RG84,Amo00,DA00} and
the momentum-space representation~\cite{Chi98}.
Nowadays, the microscopic description of nucleus-nucleus ($AA$)
scattering becomes a hot topic~\cite{Fur08}.
In all these models,
$W$ comes from the imaginary part of the effective $NN$ interaction,
for which a Br\"uckner $g$ matrix evaluated in infinite nuclear matter
is often adopted. Despite the great success of this folding-model
approach, it is quite obvious that the use of the $g$ matrix is
not feasible for low-energy scattering, because individual energy-levels
of the $N+A$ system near the nucleon threshold strongly affect the
scattering process.

An alternative approach to the microscopic description of
$NA$ elastic scattering is many-body calculation
of the $N+A$ system including channel-couplings to the continuum
states.
In this approach, a $NN$ effective interaction with no imaginary
part is used, and $W$ is generated through the
coupling to non-elastic channels that are explicitly taken into
account.
One of the most suitable models for this purpose will be
the particle-vibration coupling (PVC) method, which describes
collective vibrations and single-particle motion of individual
nucleons simultaneously.
In a recent paper~\cite{CPVC},
the microscopic continuum PVC (cPVC) method has been
proposed and applied to studies on single-particle (sp)
structures in $^{40}$Ca, $^{208}$Pb, and $^{24}$O. In general, the
sp states of a nucleus $A$ are observed in excitation energy spectra
in the neighboring $A\pm 1$ nuclei. It was shown in Ref.~\cite{CPVC}
that the cPVC method describes quite well the fragmentation
of the sp hole- and particle-states as well as the shift of those
centroid energies, in good agreement with experimental data.
These are distinguished features of sp states that are not
taken into account in the standard Hartree-Fock (HF) picture.

The cPVC method is based on the self-consistent microscopic
HF and the continuum random-phase-approximation
(RPA)~\cite{mizuyama,sagawa} with the Skyrme effective interaction.
In this framework, the microscopic nucleon optical potential
is characterized by the nucleon self-energy
corresponding to specific energy $E$ in the asymptotic region
of the $N+A$ system; $E$ can be interpreted
as the incident energy
of the nucleon on the target nucleus $A$
in the optical model picture.
A great advantage of the cPVC method to other existing PVC
models~\cite{colo10,litvinova,anatolj,Giai} is, as emphasized in
Ref.~\cite{CPVC}, the proper treatment of the continuum with
imposing an explicit boundary condition on a nucleon in a continuum state.
Therefore, it is quite promising that the cPVC method can be
applied to studies on reaction observables, for which
boundary conditions play essential roles in general.

In this Rapid Communication, we apply the cPVC method to the
neutron scattering on $^{16}$O below 30~MeV.
We see how the cPVC method can describe the
absolute values and energy dependence of $\sigma_{\rm R}$ in particular.
Important aspects of the present study, in comparison with
the preceding works~\cite{Giai,Nob11} along the same line,
are as follows. First, we treat the continuum explicitly
as mentioned above. Second, we include a very large model space
(various phonon states up to 60~MeV) as described below.
Third, we are interested in the structures, i.e., nontrivial
energy dependence, of $\sigma_{\rm R}$
due to the PVC, expected to appear at quite lower incident energies.

In the cPVC framework,
the scattering wave function
of neutron $\Psi_{\mathrm{PVC}}(\boldsymbol{r}\sigma,\boldsymbol{k})$
from $A$, with the relative coordinate $\boldsymbol{r}$, the
intrinsic coordinate $\sigma$ due to the spin degrees of freedom,
and the relative wave number $\boldsymbol{k}$ in the asymptotic region,
is described by the following
Lippmann-Schwinger equation
\begin{eqnarray}
\Psi_{\mathrm{PVC}}^{\left(  +\right)  }\left(  \boldsymbol{r}\sigma
,\boldsymbol{k}\right)
&=&
\phi_{\mathrm{F}}\left(  \boldsymbol{r}%
\sigma,\boldsymbol{k}\right)
\nonumber \\
&&+\sum_{\sigma^{\prime}\sigma^{\prime\prime}}%
\int\int d\boldsymbol{r}^{\prime}d\boldsymbol{r}^{\prime\prime}
G^{\left(  +\right)}
\left(  \boldsymbol{r}\sigma,\boldsymbol{r}^{\prime}\sigma^{\prime};E\right)
\nonumber \\
&&\;\;\times
\left[  v\left(  \boldsymbol{r}^{\prime}\sigma^{\prime}\right)  \delta\left(
\boldsymbol{r}^{\prime}-\boldsymbol{r}^{\prime\prime}\right)  \delta
_{\sigma^{\prime}\sigma^{\prime\prime}}
\right.
\nonumber \\
&&\left.\;\;\;\;\;\;
+\Sigma\left(  \boldsymbol{r}^{\prime}
\sigma^{\prime},\boldsymbol{r}^{\prime\prime}\sigma^{\prime\prime};E\right)
\right]  \phi_{\mathrm{F}}\left(  \boldsymbol{r}^{\prime\prime}\sigma^{\prime\prime},\boldsymbol{k}\right),
\nonumber\\
\label{LSeq}
\end{eqnarray}%
where $\phi_{\mathrm{F}}$ denotes the neutron free wave
and $v\left(  \boldsymbol{r}^{\prime}\sigma^{\prime}\right)$ is the
HF one-body mean-field potential.
The PVC Green function and
the corresponding self-energy are denoted by
$G^{\left(  +\right)}
\left(  \boldsymbol{r}\sigma,\boldsymbol{r}^{\prime}\sigma^{\prime};E\right)$
and
$\Sigma\left(  \boldsymbol{r}^{\prime}
\sigma^{\prime},\boldsymbol{r}^{\prime\prime}\sigma^{\prime\prime};E\right)$,
respectively.
With the partial wave expansion,
one may find that the transition matrix ($T$~matrix)
of the elastic scattering is given by
\begin{eqnarray}
T_{lj}^{\mathrm{PVC}}\left(  E\right)
&=&
\lim_{r\rightarrow\infty}\frac
{2i}{rh_{l}^{(+)}\left(  kr\right)  }
\int\int dr^{\prime}%
dr^{\prime\prime}
r^{\prime\prime}
G_{lj}^{\left(  +\right)  }\left(  rr^{\prime};E\right)
\nonumber \\
&&
\hspace*{-3mm}\times
\left[
v_{lj}\left(  r^{\prime}\right)
\delta(r'-r'')
+\Sigma_{lj}\left(  r^{\prime}r^{\prime\prime};E\right)
\right]j_{l}\left(  kr^{\prime\prime}\right),
\nonumber \\
\label{Tmat}
\end{eqnarray}
where $l$ ($j$) represents the orbital angular momentum
(total single-particle spin) of neutron, and
$h_l^{(+)} (kr)$  and $j_l (kr)$ are, respectively,
the spherical Hankel function with the outgoing asymptotics
and the spherical Bessel function.
Explicit expressions of
$G^{\left(  +\right)}$ and $\Sigma$ are given by
Eqs.~(6) and (7) of Ref.~\cite{CPVC}, respectively.
Note that we use $E$ for the relative energy between neutron and $A$,
which is denoted by $\omega$ in Ref.~\cite{CPVC}. Furthermore,
we put $(+)$ in the superscript of $\Psi_{\rm PVC}$, $G$, and
$G_{lj}$, to explicitly represent that these functions satisfy
the outgoing boundary condition.
We solve a Dyson equation, Eq.~(10) of Ref.~\cite{CPVC}, to
obtain $G_{lj}$, and hence $T_{lj}^{\rm PVC}$. All other details
can be found in Ref.~\cite{CPVC}.

The differential elastic cross section $d\sigma/d\Omega$ is
given by
\beq
\frac{d\sigma}{d\Omega}
=
|\mathcal{F}(\theta)|^2
+
|\mathcal{G}(\theta)|^2,
\label{diffcross}
\eeq
where
\bea
\mathcal{F}(\theta)
&=&
\frac{1}{2ik}
\sum_{lj}
\frac{2j+1}{2}
\left\{-iT_{lj}(E)\right\}
P_l(\cos\theta),
\\
\mathcal{G}(\theta)
&=&
\frac{\sin\theta}{2k}
\sum_{l(\neq 0)j}
\frac{2j+1}{2}
\frac{j(j+1)-l(l+1)-3/4}{l(l+1)}
\nonumber\\
&&\times
\left\{-iT_{lj}(E)\right\}
P'_l(\cos\theta).
\eea
Here, $P_l$ and $P_l'$ are, respectively, the Legendre polynomial and
its derivative with respect to $\cos\theta$.
The total cross section $\sigma_{\rm tot}$ and the total-elastic cross
section $\sigma_{\rm el}$ are given by
\begin{eqnarray}
\sigma_{\rm tot}(E)
&=&
\sum_{lj}
\frac{2\pi}{k^2}
\frac{2j+1}{2}
\left[\text{Im}\hspace{0.1cm}T_{lj}(E)\right]
\equiv
\sum_{lj}\sigma_{{\rm tot;}lj}(E),
\nonumber \\
\label{totcross}
\\
\sigma_{\rm el}(E)
&=&\sum_{lj}
\frac{\pi}{k^2}
\frac{2j+1}{2}
|T_{lj}(E)|^2
\equiv
\sum_{lj}
\sigma_{{\rm el;}lj}(E).
\label{totcrossel}
\end{eqnarray}
The reaction cross section $\sigma_{\rm R}$ is defined by
\beq
\sigma_{\rm R}(E)=
\sigma_{\rm tot}(E)-\sigma_{\rm el}(E).
\eeq

As one may see from Eqs.~(\ref{LSeq})--(\ref{Tmat}), the self-energy
$\Sigma$ serves as a dynamical polarization potential. Therefore,
the imaginary part $W$ of the optical potential comes from
$\Sigma$, which takes into
account the coupling within the Skyrme continuum-RPA response function
and the HF Green function~\cite{CPVC}.
Note that we explicitly treat the nonlocality of $\Sigma$,
whereas it was approximately localized in the previous work~\cite{Giai}.
The HF $T$~matrix can be obtained by
\begin{eqnarray}
T_{lj}^{\mathrm{HF}}\left(  E\right)
&=&
\lim_{r\rightarrow\infty}\frac
{2i}{rh_{l}^{(+)}\left(  kr\right)  }
\int dr^{\prime}%
r^{\prime}
G_{0,lj}^{\left(  +\right)  }\left(  rr^{\prime};E\right)
\nonumber \\
&&
\times
v_{lj}\left(  r^{\prime}\right)
j_{l}\left(  kr^{\prime}\right),
\label{TmatHF}
\end{eqnarray}
where $G_{0,lj}^{\left(  +\right)}$ means the HF Green function.

In the present calculation, we adopt
the Skyrme $NN$ effective interaction SkM*~\cite{SKMs}.
Note that, in contrast to in Ref.~\cite{Nob11}, we do not
introduce any other interactions to the calculation of scattering
observables, i.e., a fully consistent treatment of
the effective interaction is carried out.
For the cPVC calculation, as in Ref.~\cite{CPVC},
the orbital angular momentum cutoff for the unoccupied continuum states
is set at $l_{\rm cut}=7\hbar$,
and we include phonons associated with the multipolarities
$\lambda^\pi$
of $2^+,3^-,4^+$, and $5^-$, up to 60~MeV of the RPA excitation energy.
We choose the Fermi momentum $k_{\rm F}=1.33$~fm$^{-1}$ for
the residual force in the self-energy function of the PVC calculation.
The radial mesh size is $\Delta r=0.2$~fm and the maximum value of
$r$ is set to 20~fm, to obtain the $T$~matrix by Eq.~(\ref{Tmat}).

%
\begin{figure}[htpb]
\begin{center}
\includegraphics[width=0.35\textwidth,angle=-90,clip]{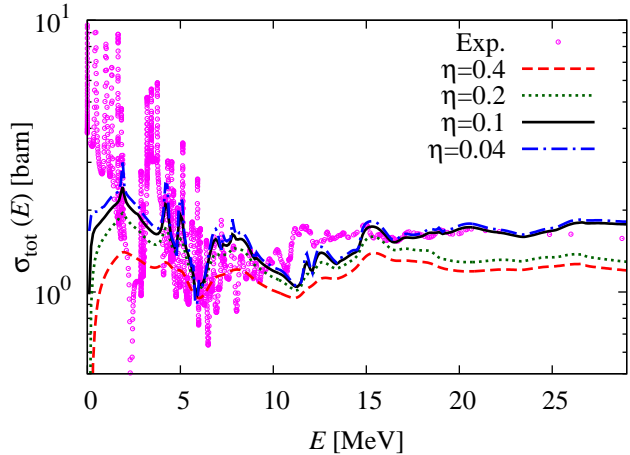}
\caption{(Color online)
Total cross section of neutron scattering by $^{16}$O,
as a function of the incident energy.
The dashed, dotted, solid, and dash-dotted lines
show the results with $\eta=0.4$, 0.2, 0.1, and 0.04~MeV, respectively.
Experimental data are taken from Ref.~\cite{EXPdata}.
}
\label{sigtot}
\end{center}
\end{figure}
First, we show in Fig.~\ref{sigtot} the $\eta$-dependence
of $\sigma_{\rm tot}(E)$ of neutron scattering by $^{16}$O;
$\eta$ is a parameter introduced in the evaluation of the Green
function and the RPA response function~\cite{CPVC}, and corresponds to
the resolution scale of $E$.
The dashed, dotted, solid, and dash-dotted lines
show the results with $\eta=0.4$, 0.2, 0.1, and 0.04~MeV, respectively.
Considering the experimental situation, $E$ should be an order of eV,
which cannot be achieved because of the computational limitation.
As one sees from the figure, however, $\sigma_{\rm tot}(E)$ for
$E\ge4.0$~MeV converges at $\eta=0.1$~MeV. Thus, in the following
discussion we use $\eta=0.1$~MeV and restrict ourselves for $E\ge4.0$~MeV.

In Fig.~\ref{sigtot},
the solid line reproduces quite well the energy dependence of
the experimental data~\cite{EXPdata} except for $10\la E \la 15$~MeV.
A remarkable feature is the reproduction of the fine structure of
$\sigma_{\rm tot}(E)$ in part, which is the benefit of
the PVC. It is well known that many peaks in the experimental
$\sigma_{\rm tot}(E)$ correspond to the compound resonance.
It is quite obvious that the present cPVC calculation cannot describe
such peaks. Rather than that, the cPVC method is considered to well
describe the so-called doorway states~\cite{Wei61}. Keeping this in mind,
we see that the present calculation reasonably reproduces the
energy dependence of $\sigma_{\rm tot}(E)$. Note that we do not
tune any adjustable parameters, although the result can, to some extent,
depend on the Skyrme parameters adopted.

%
\begin{figure}[htpb]
\begin{center}
\includegraphics[width=0.35\textwidth,angle=-90,clip]{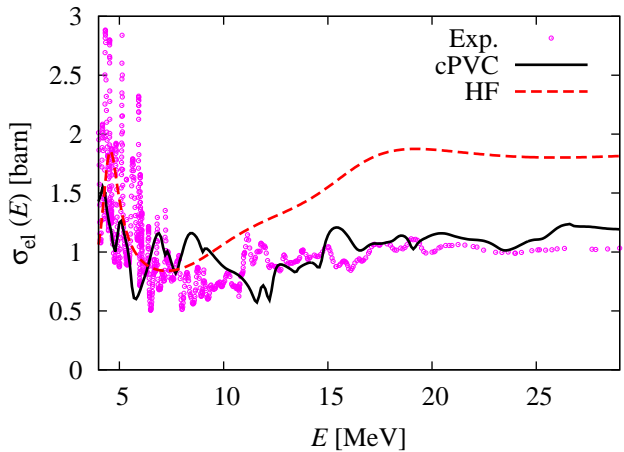}
\caption{(Color online)
Total-elastic cross section of the neutron scattering by $^{16}$O.
The solid (dashed) line shows the result of the cPVC method
(HF calculation).
Experimental data are taken from Ref.~\cite{EXPdata}.
}
\label{sigelas}
\end{center}
\end{figure}
Figure~\ref{sigelas} shows the $\sigma_{\rm el}(E)$ calculated by
the cPVC method (solid line) and the HF calculation (dashed line).
Results are plotted in the linear scale.
The cPVC method
reproduces well the experimental data~\cite{EXPdata},
whereas the HF calculation overshoots them by about 80\%
above 15~MeV. This shortcoming of the HF calculation is
rather trivial because of the absence of the imaginary
part of the optical potential;
absorption effects become more significant
as $E$, the number of open channels in fact, increases.
It should be noted that the HF calculation
gives $\sigma_{\rm el}(E)=\sigma_{\rm tot}(E)$ and hence
$\sigma_{\rm R}(E)=0$.
Another important finding is that the HF result has only one peak
around 4.6~MeV, which turns out to be due to the ${\rm f}_{7/2}$ sp orbit.
The cPVC method gives much more complicated shapes of
$\sigma_{\rm el}(E)$, as in Fig.~\ref{sigtot}. This
result clearly shows the importance of the PVC in the neutron
elastic scattering at low energies.

%
\begin{figure}[htpb]
\begin{center}
\includegraphics[width=0.35\textwidth,angle=-90,clip]{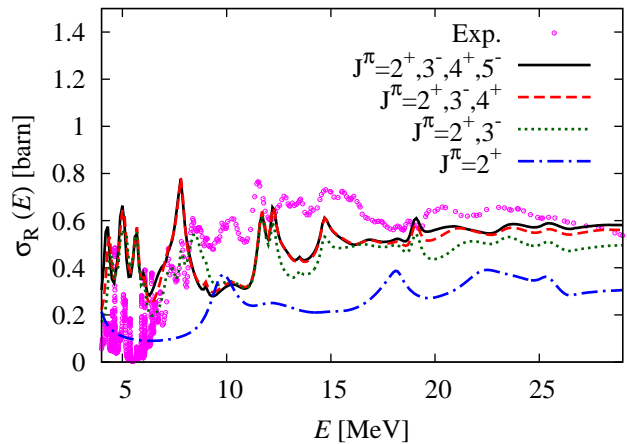}
\caption{(Color online)
Reaction cross section of neutron for $^{16}$O.
The solid, dashed, dotted, and dash-dotted lines correspond to
the calculation with $\lambda_{\rm max}=5$, 4, 3, and 2, respectively.
Experimental data are taken from Ref.~\cite{EXPdata}.
}
\label{sigr}
\end{center}
\end{figure}
Next we compare the result of $\sigma_{\rm R}(E)$ (solid line) with
the experimental data~\cite{EXPdata} in Fig.~\ref{sigr}.
Again, we plot the result in the linear scale. Although the theoretical
$\sigma_{\rm R}(E)$ slightly undershoots (overshoots)
the data for $10 \la E \la 25$~MeV ($E \la 6$~MeV),
some peak structure is reproduced well.
On the average for $ 8 \la E \la 30$~MeV,
the discrepancy between the theoretical
values and experimental data for $\sigma_{\rm R}(E)$ is about 15\%.
It should be noted that the present cPVC calculation describes the
$\sigma_{\rm R}(E)$ only though channel-coupling effects,
i.e., with no imaginary part of an effective interaction.
It will be a remarkable achievement
that about 85\% of the $\sigma_{\rm R}(E)$ is successfully explained in
this manner. It was shown in Ref.~\cite{Nob11} that more than half the
$\sigma_{\rm R}(E)$ for proton scattering on $^{58}$Ni, $^{48}$Ca, and
$^{90}$Zr is due to a coupling with the deuteron (transfer) channel.
In the present calculation, the deuteron channel seems to play less
important roles; it should be remarked that i) theoretical
calculation shows good agreement with data for $E\ga 25$~MeV
and ii) we still have undershooting around 10~MeV, where
the $(n,d)$ reaction channel is closed.

In Fig.~\ref{sigr}, the dependence of $\sigma_{\rm R}(E)$
on the maximum multipolarity $\lambda_{\rm max}$ is also shown.
The dashed, dotted, and dash-dotted lines correspond to
the calculation with $\lambda_{\rm max}=4$, 3, and 2, respectively;
the solid line represents the full calculation with $\lambda_{\rm max}=5$.
One sees a good convergence at $\lambda_{\rm max}=4$.
Another finding is the role of $\lambda=3$, which essentially
generates the many peaks in $\sigma_{\rm R}(E)$;
$\lambda=4$ then gives a slight change in
the shapes and positions of the peaks.

%
\begin{figure}[htpb]
\begin{center}
\includegraphics[width=0.35\textwidth,angle=-90,clip]{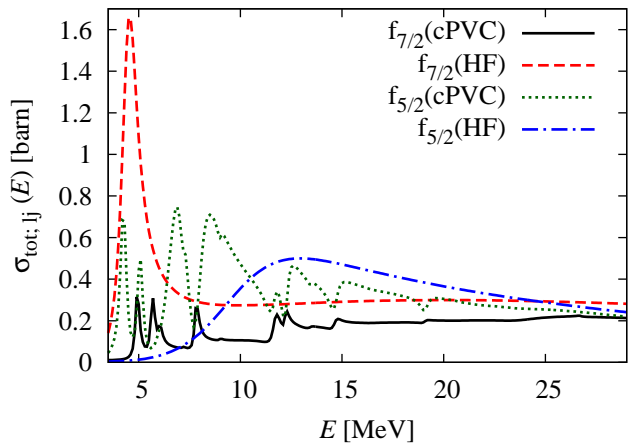}
\caption{(Color online)
Partial cross sections of $\sigma_{\rm tot}$.
The solid (dashed) and dotted (dash-dotted) lines represent
the results of the ${\rm f}_{7/2}$ and ${\rm f}_{5/2}$ orbits
obtained by the cPVC method (HF calculation).
}
\label{pxtot}
\end{center}
\end{figure}
To see how the PVC gives peaks in the cross section more clearly,
we show in Fig.~\ref{pxtot} the partial cross sections of
$\sigma_{\rm tot}(E)$,
$\sigma_{{\rm tot;}lj}(E)$ in Eq.~(\ref{totcross}), for the f-orbits.
The cPVC (HF) result for the ${\rm f}_{7/2}$ and ${\rm f}_{5/2}$ orbits
are shown by the solid (dashed) and dotted (dash-dotted) lines, respectively.
The dashed line has a resonant peak around 4.6~MeV, which is fragmented
by the PVC as shown by the solid line. Similarly, the PVC changes
the dash-dotted line
to the dotted line, generating nontrivial structures. Thus, the PVC completely
changes the shapes of the partial cross sections. Fragmentation of the sp strength
function is necessary to explain the energy dependence of
experimental data, as shown in Figs.~\ref{sigtot}--\ref{sigr}.

%
\begin{figure}[htpb]
\begin{center}
\includegraphics[width=0.34\textwidth,angle=-90,clip]{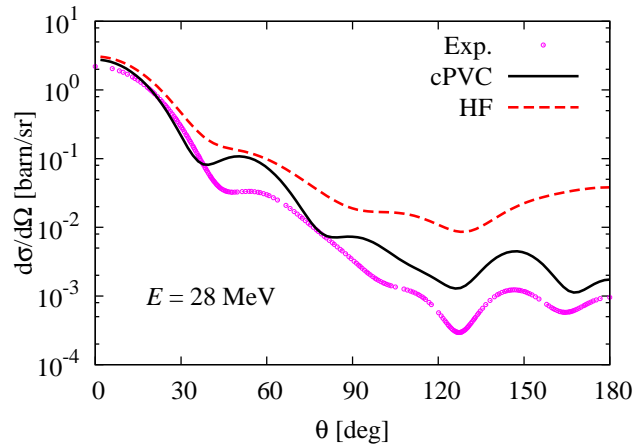}
\caption{(Color online)
Angular distribution of the neutron elastic cross section
by $^{16}$O at 28~MeV.
The solid (dashed) line shows the result of the cPVC method (HF calculation).
Experimental data are taken from Ref.~\cite{EXPdata}.
}
\label{angdis}
\end{center}
\end{figure}
Finally, we show the result of the
differential cross section $d\sigma/d\Omega$ in Fig.~\ref{angdis}.
The cPVC and HF results are denoted by the solid and dashed lines, respectively.
The dashed line severely overshoots the data at backward angles. This issue remains
for the solid line, although the agreement with the experimental data
is significantly improved by including the PVC.
This may indicate that the radial dependence of the optical potential is not
correctly generated by the present framework; further investigation on
this is our future work.

In this study, we applied the continuum particle-vibration coupling (cPVC) method
to the calculation of scattering observables,
i.e., the total ($\sigma_{\rm tot}$), total-elastic ($\sigma_{\rm el}$),
and reaction cross sections ($\sigma_{\rm R}$),
of neutron on the $^{16}$O target at 4--30~MeV.
The cPVC method describes the single-particle motion and collective vibrations
simultaneously, with properly treating the boundary condition of a nucleon in
continuum states. This feature makes the calculation of the
transition matrix ($T$~matrix) straightforward.
We used the Skyrme interaction (SkM*) as an effective nucleon-nucleon ($NN$) interaction,
and included phonon states of $^{16}$O up to
the $5^-$ state with excitation energy
of 60~MeV. The $T$~matrix is evaluated with the PVC Green function, which is
obtained by solving the Dyson equation. The couplings to various collective
states are characterized by the self-energy $\Sigma$. Since the cPVC method employs
a real $NN$ interaction, the imaginary part of the optical potential
purely comes from
$\Sigma$. In the present framework, $\sigma_{\rm R}$
is described as a loss of the incident flux to various channels
that are explicitly taken into account.

The results of the present cPVC calculation satisfactorily agree
well with experimental data of $\sigma_{\rm tot}$ and $\sigma_{\rm el}$.
For $\sigma_{\rm R}$, the cPVC method explains
about 85\% of the experimental data on average, which will be an important
achievement of the mean-field type calculation for neutron scattering.
Another remarkable feature of the cPVC result
is the fragmentation of a single-particle resonant cross section.
This results in a good correspondence with some of the peaks
observed, probably those due to the doorway states.
On the other hand, the Hartree-Fock
(HF) calculation was found to give a rather trivial shape of the cross section
and severely overshoot $\sigma_{\rm el}$ for $E\ga 10$~MeV.
The cPVC method describes the angular distribution of the elastic cross
section
much better than the HF calculation. However, there still remains a
discrepancy at scattering angles larger than $30^\circ$.
This will be due to an incorrect radial dependence of the optical potential
in the cPVC framework.

In future, we will apply the cPVC method to other reaction systems,
including proton scattering. Application to inelastic scattering and
photo-induced reactions will also be important.
Another important subject is the interaction dependence. The scattering
observables can be used as new constraints on the parameters of
the Skyrme interaction, determined to reproduce nuclear bound-state properties.
In addition to that, as discussed in Ref.~\cite{CPVC}, further investigation on the
treatment of Pauli's principle in the cPVC framework will be necessary.

The authors thank G.~Col\`o, M.~Yahiro, and K.~Hagino for helpful discussions.

\end{document}